\documentclass[preprint,amsmath,amssymb]{revtex4}          

\usepackage{graphicx}
\usepackage{dcolumn}
\usepackage{bm}

\begin{document}


\title{Putting the ``Warp'' into Warp Drive}

\author{Richard K Obousy}
\author{Gerald Cleaver}

\email{Richard\_K\_Obousy@baylor.edu}
\affiliation{%
Baylor University, Waco, Texas, 76706, USA
}%

\date{\today}

\begin{abstract}
Over the last decade, there has been a respectable level of scientific interest regarding the concept of a `warp drive'. This is a hypothetical propulsion device that could theoretically circumvent the traditional limitations of special relativity which restricts spacecraft to sub-light velocities. Any breakthrough in this field would revolutionize space exploration and open the doorway to interstellar travel. This article discusses a novel approach to generating the `warp bubble' necessary for such propulsion; the mathematical details of this theory are discussed in an article published in the Journal of the British Interpanetary Society  \cite{oc}. The theory is based on some of the exciting predictions coming out of string theory and it is the aim of this article to introduce the warp drive idea from a non-mathematical perspective that should be accessible to a wide range of readers.\end{abstract}

\maketitle

\section{Why Warp Drive?}

$\indent$The universe is truly vast, and current propulsion technology severely restricts us to the exploration of our own solar system. If we wanted to visit even the nearest star systems, we would be faced with transit times of many tens of thousands of years at best. A compelling reason for why we might actually want to visit other stars is the recent evidence of `extrasolar planets,' which are planets that orbit stars other than our sun. To date, we know of at least 250 extrasolar planets. Even more exciting is the possibility that some of these planets may be `Earth-like', with the theoretical capability to support life.

$\indent$This year, a Swiss team discovered a planet designated Gliese 581c; this planet orbits the star Gliese, 20.4 light years away from earth. The planet is remarkable in that it is the only known extrasolar planet so far that exists in the area known as the `habitable zone' of a star (Fig.1). This is the area surrounding a star where surface temperatures could maintain water in a liquid state. Gliese 581c is believed to be roughly 5 times the mass of Earth, and to have a similar surface temperature to that of Earth.


\begin{figure}[!ht]
\includegraphics[width=300pt]{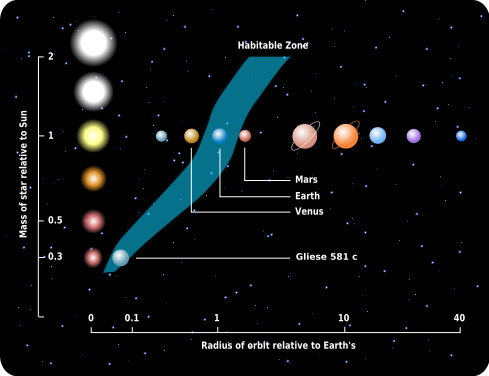}
\caption{\textit{Gliese lies within the 'Habitable Zone' of its host star, meaning that liquid water could exist on its surface.}}
\end{figure}


$\indent$Given these striking similarities, one naturally wonders if Gliese 581c might harbor life. Although it may be indirectly possible to verify the existence of life using observational techniques here on Earth, the depth of analysis would be particularly limited. On the other hand, we could obtain data of much more scientific value if we could actually visit these planets with probes, or even humans, to better understand the origin and development of life, and to see if it exists only in the form of elementary organisms or if indeed intelligence has evolved. The discovery of life outside of Earth would obviously be of huge scientific significance, and could quite possibly be remembered historically as the most significant scientific discovery of all time. The only way we could realistically visit these worlds in time-frames on the order of a human lifespan would be to develop what has been popularly termed a `warp drive'.

\section{Origins of the Warp Drive}
$\indent$
The term `warp drive' originated in science fiction. A 1994 paper by theoretical physicist Miguel Alcubierre placed the concept on a more scientific foundation \cite{alc}. Alcubierre's paper demonstrated that a solution to Einstein's field equations could `stretch' space in a way such that space itself would expand behind a hypothetical spacecraft, while contracting in front of the craft, creating the effect of motion (Fig. 2). In contrast to the conventional technology that results in movement of the craft through space, in this theory space itself moves around the spacecraft. This is a radical departure from the traditional concept of motion, because the spacecraft is, in a classical sense, motionless within a hypothetical bubble of transient spacetime.


\begin{figure}[!ht]
\includegraphics[trim = 20mm 30mm 20mm 10mm, clip,width=220pt]{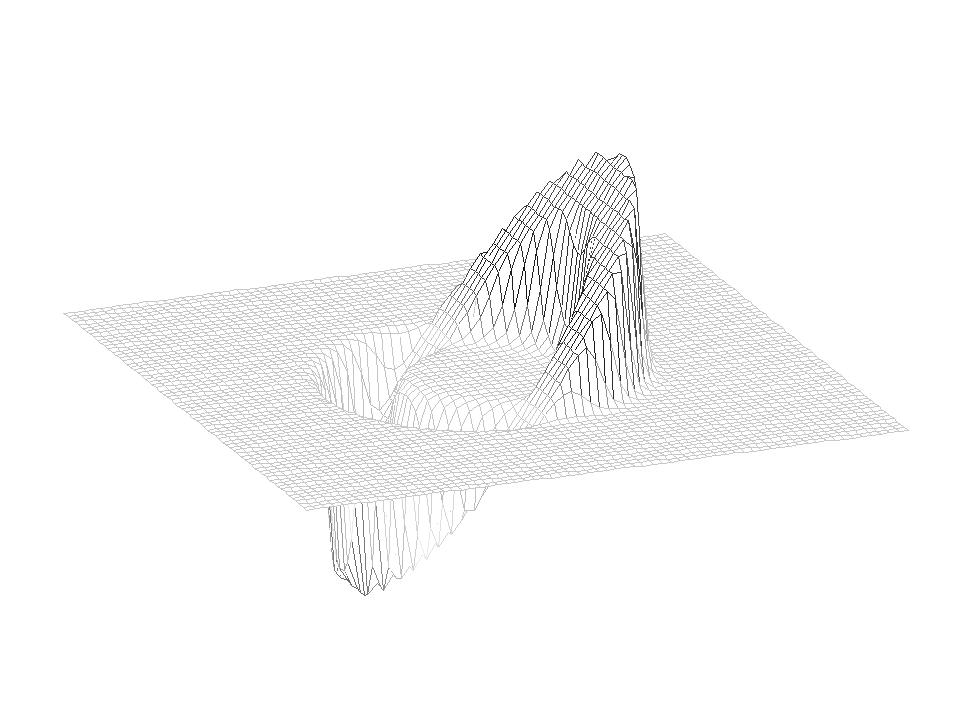}
\caption{\textit{The Alcubierre `top-hat' metric. A bubble of assymetric spacetime curvature surrounds a spacecraft which would sit in the center of the bubble.}}
\end{figure}


$\indent$What is particularly appealing about this approach to propulsion is that the spacecraft could, in theory, effectively travel faster than the speed of light. Special relativity forbids objects from moving through space at or above the speed of light, but the fabric of space itself is not restricted in any way. Thus, even though the spacecraft cannot travel faster than light in a local sense, it could make a round trip between two points in an arbitrarily short period of time as measured by an observer who remained at rest at the starting point.

\section{Elementary Physics of Warp Drives}

$\indent$The warp drive concept is based on Einstein's General theory of relativity (GR), an accepted and extremely well-tested physical theory. One of the necessary components of a warp drive is an exotic form of energy called negative energy, which would have to be produced in copious amounts for propulsion of a spacecraft to occur (See Fig. 3). Negative energy is classically forbidden in general relativity; however, negative energy is allowed if we shift into the paradigm of quantum field theory. And actually, negative energy is not only allowed in this theory; it has been experimentally verified using the `Casimir Effect'\cite{cas}. 

\begin{figure}[!ht]
\includegraphics[trim = 20mm 40mm 20mm 10mm, clip,width=320pt]{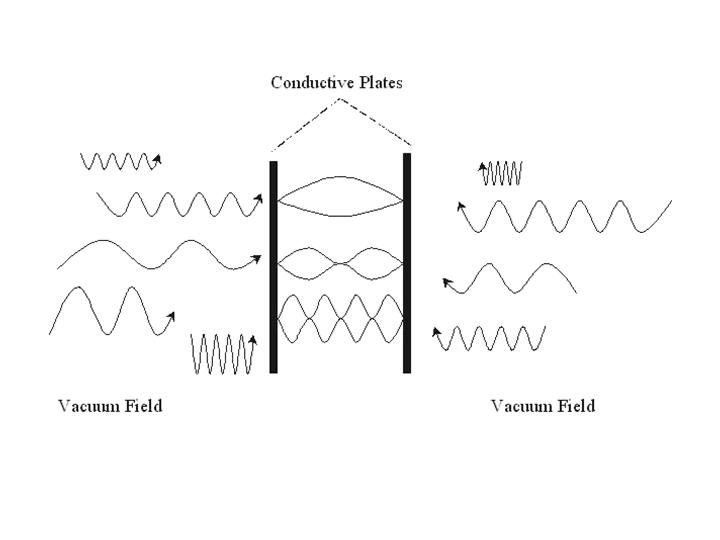}
\caption{\textit{External to the conducting plates there is no restriction on the modes on the vacuum. On the interior only standing waves may form. This asymmetry in the field causes the plates to be pulled together by a force that is purely quantum in nature. }}
\end{figure}

$\indent$
The Casimir Effect is one of the most salient manifestations of the vacuum fluctuations in existence. Its simplest realization is found in the interaction between a pair of neutral, parallel conducting plates. The plates modify the ground state of the quantum vacuum in the volume between the plates, creating an attractive force between the plates. For a good review see \cite{m}.

$\indent$The physical interpretation is that a negative energy state exists in the interior region of the plates. Theoretically, the Casimir Effect could be used to create the negative energy required for a warp drive. From this perspective, there is nothing that theoretically prevents the creation of warp drive.

\section{Nature as Teacher}

$\indent$In essence, the warp bubble consists of a region of expanding spacetime, and a region of contracting spacetime. The first stage in developing a working warp drive necessarily involves learning how to generate this asymmetric bubble.

$\indent$As in all good engineering technologies, nature herself can provide a degree of insight. We are fortunate in that spacetime is already expanding, albeit at an extremely slow rate. In 1929, Hubble's observation of galactic redshifting cemented the paradigm of an expanding spacetime in physical cosmology (Fig. 4). 

\begin{figure}[!ht]
\includegraphics[width=160pt]{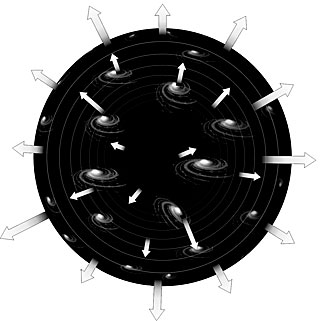}
\caption{\textit{Seen from Earth most galaxies appear to be receeding from us. This provides us with evidence that the universe is expanding.}}
\end{figure}

$\indent$The energy responsible for the current expansion of spacetime is generically termed the `cosmological constant', or equivalently, the `vacuum energy' (the terms will be used interchangeably in this article). On a local scale, space is expanding extremely slowly: around a billion billionth of a meter per second per meter. Clearly, to build a warp drive would require spacetime to be stimulated in some way to expand (and contract) at a much higher rate, but the fact that it is already expanding gives us a primitive roadmap to realizing the ambitions of warp drive. The first step to controlling a mechanism is to understand the mechanism; thus if we can understand why spacetime is already expanding, we may be able to use this knowledge to artificially generate an expansion (and contraction) of spacetime.

\section{Origins of the Cosmological Constant}

$\indent$Understanding the cosmological constant may play a crucial role in the development of warp drive, as it will help us to understand why space is expanding. However, the origin of the cosmological constant is still a mystery nearly a century after its introduction into cosmology \cite{w}. Physicists are not certain what generates the cosmological constant; we simply know that it is there. Several ideas exist as to the nature of this field \cite{bdel} \cite{bv} \cite{cg} \cite{c} \cite{cr} \cite{a} \cite{e} \cite{gpt}. Dark energy, for example, is a popular contemporary phrase. Efforts have been made to explain the cosmological constant using quantum field theory (QFT), which was created several decades after Einstein's general relativity (GR). However, theoretical predictions using QFT are in such huge conflict with observations, being off by a factor of $10^{120}$,  that the prediction has been called``the worst prediction of theoretical physics.''  

$\indent$One (partial) fix to the vacuum energy calculations is the introduction of supersymmetry (SUSY). A full discussion of SUSY is beyond the scope of this article; however, the basic idea is that all known particles have an associated superparticle whose `spin' differs by half a unit of spin. Thus, in SUSY all the integer spin particles (bosons) are balanced by half-integer spin particles (fermions) in the elementary particle content of the Universe.

$\indent$This idea is particularly appealing in that the existence of supersymmetric particles tames the embarrassing predictions of QFT with regards to the vacuum energy calculations. In our model analyze the contributions to the vacuum energy density that are purely higher dimensional in origin.
\section{Physics in Higher Dimensions}

$\indent$The idea of extra dimensions is not a new one. It was first introduced by the physicist Theodore Kaluza in 1919, in an effort to unify the laws of gravitation and electromagnetism \cite{K}. Kaluza's main theory was to introduce a 5th dimension. Upon solving Einstein's equations (or more technically, varying the Einstein-Hilbert action), Kaluza was able to produce a set of equations which contained both the Einstein field equations and Maxwell's field equations. The question of where the extra dimension was located was addressed by Oskar Klein who, in 1926, suggested that the fifth dimension compactifies so as to have the geometry of a circle of extremely small radius \cite{Kln}. One way to envisage this spacetime is to imagine a hosepipe. From a long distance it looks like a one dimensional line but a closer inspection reveals that every point on the line is in fact a circle (Fig. 5). Although the theory did contain flaws, it was the first hint that extra-spatial dimensions may play an important role in physics.

\begin{figure}[!ht]
\includegraphics[trim = 00mm 40mm 0mm 00mm, clip,width=220pt]{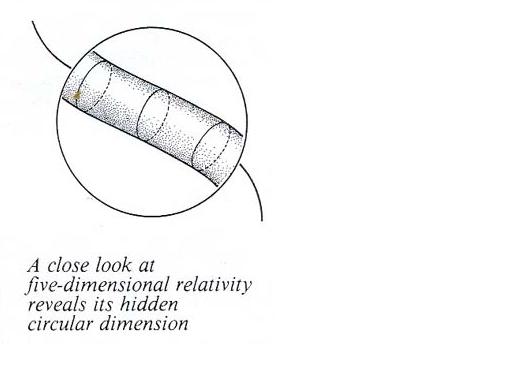}
\caption{\textit{An extra dimension remains hidden to us due to its size.}}
\end{figure}

$\indent$ More recently, extra dimensions have become an accepted part of modern theoretical physics. Superstring theory, or M-theory in its most modern guise, is a theory that attempts to unify all known physics under a single mathematical and conceptual framework, and predicts the existence of extra spatial dimensions. Two other contemporary extra-dimensional models are the Randall-Sundrum (RS) \cite{rs1} model of `warped' extra dimensions, and the Arkani-Hamed-Dimopoulos-Dvali (ADD) \cite{add} model of large extra dimensions. Both theories attempt to explain the observation that gravity is far weaker than the other known forces. One way to think about the ADD model is to picture gravity as being free to propagate in all dimensions (including the extra ones), while other forces are restricted to our familiar three spatial dimensions. Thus, gravity is, in a sense, diluted, which results in its being much weaker than the other forces.


\section{Manipulating Extra Dimensions to Create a Warp Drive} 

$\indent$There are currently numerous models which attempt to explain the physical origin of the cosmological constant. One model suggests that the zero-point energy of graviton fluctuations in the extra dimensions of the ADD model may ultimately be responsible \cite{g}. The zero point energy is precisely the energy that causes the plates in the Casimir effect to attract one another, as discussed earlier. This zero-point energy should also exist in higher dimensions. In this picture, it is the periodicity of the extra dimension that plays the exact same role as the parallel plates in the traditional Casimir effect. Just as the plates ensure that deconstructive interference destroys non-resonant frequencies of the quantum vacuum, so too does a circular extra dimension, which allows only frequencies that are resonant along its circumference. In this model it is the size of the extra dimension that directly regulates that magnitude of the cosmological constant, and therefore the expansion of spacetime. 

$\indent$In a recent  paper \cite{oc}, we addressed the plausibility of locally influencing the size of the extra dimension to locally (by local, we mean in the vicinity of a spacecraft) adjust the cosmological constant. This could theoretically create a modification of spacetime around a craft that could be tuned to acquire the characteristics of the Alcubierre bubble (see Figure 3). The basic idea is that by altering the radius of an extra dimension, it would be possible, in principle, to adjust the energy density of spacetime (which relates directly to the cosmological constant which ultimately controls the inflation/contraction of space itself). We have taken two approaches to this concept: one from the viewpoint of QFT another from GR. The equations of both theories indicated that the physics of the extra dimensional space effects the expansion rate of `normal' space by a `dimensional shearing' effect. The equations of GR demonstrated that shrinking the extra dimension would inflate our space \cite{L}, and that expanding the extra dimension would contract our space. In this way, a bubble of expanding/contracting spacetime could be created at the rear/front of a spacecraft(Fig. 6).
\begin{figure}[!ht]
\includegraphics[width=300pt]{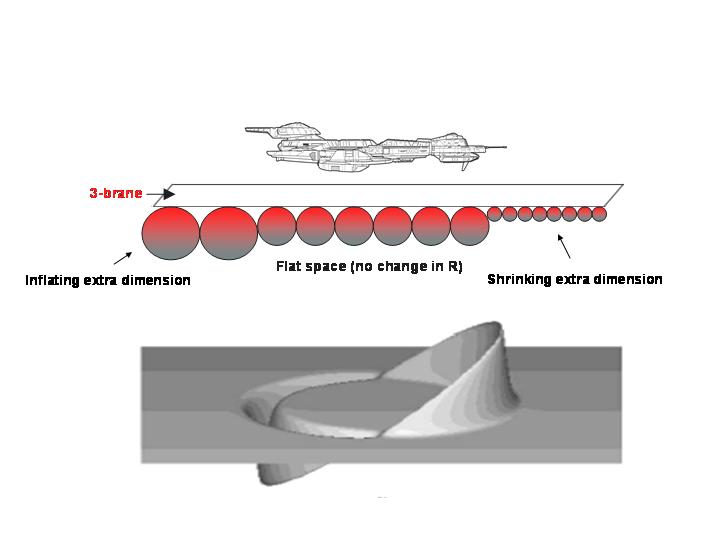}
\caption{The Alcubierre metric (lower image) is an asymmetric distortion of spacetime. A hypothetical spacecraft sits inside the bubble while space is expanded behind the spacecraft and contracted in front. This effect is achieved by manipulating the extra dimensional radius as illustrated in top image.}\end{figure}

$\indent$ Early calculations indicated that superluminal propulsion for a ship of volume $1000$m$^3$ could be achieved at an estimated energy cost of $10^{45}J$, or roughly the total mass-energy contained within the planet Jupiter after using the famous relation $E=mc^2$. Although this number may appear enormous, it is certainly an improvement on earlier calculations, which indicated that the warp drive would require more mass-energy than is contained in the entire observable universe.
$\indent$QFT calculations revealed an upper bound on the velocity a warp drive might obtain, resulting from a minimum length for an extra dimension, the Planck length. If the extra dimension were shrunk to the Planck length, then our calculations reveal the limit on warp drive velocity to be $10^{32}c$ (where c is the speed of light). This number is a theoretical bound, as our calculations regarding the energy required to reach this velocity indicate that significantly more mass-energy would be required than is available in the observable universe.

$\indent$One issue that needs to be addressed is the discrepancy between the predictions of GR regarding the expansion and contraction of spacetime and the predictions of QFT. In the limit of flat spacetime with zero cosmological constant, GR shows that the shear of a $\textit{contracting}$ extra dimension has the effect of $\textit{expanding}$ the remaining dimensions and similiarly an $\textit{expanding}$ extra dimension will $\textit{contract}$ the remaining dimensions. However, quantum field theoretic calculations show that a $\textit{fixed}$ compactification radius can also result in an expansion of the three-volume due to the Casimir effect. Although this issue clearly needs to be resolved, both theories suggest that the physics of the compactified space affect the non-compact space.

\section{Future Outlook}

$\indent$A new approach to generating the warp bubble necessary for warp drive has been proposed; this warp bubble would theoretically allow a spacecraft to travel at arbitrarily high velocities. One vital aspect of future research in this area would involve studying how to locally manipulate an extra dimension. String theory suggests that dimensions are globally held compact by strings wrapping around them; if this is indeed the case, then it may be possible to locally increase or decrease the string tension, or even counter the effects of some string winding modes. This would achieve the desired effect of changing the size of the extra dimensions, which would theoretically lead to propulsion at greater than lightspeed. 

$\indent$This approach, although highly theoretical at this stage, gives us a glimpse as to how one might address the problems associated with the vast distances involved in interstellar travel, and also opens up exciting new avenues for future research.

\section{Acknowledgements}

R.O. would like to thank Zurab Silagadze for helpful comments that improved the readability of this article.

\end{document}